# A review of the optical properties of alloys and intermetallics for plasmonics

M G Blaber, M D Arnold and M J Ford*
Institute for Nanoscale Technology, Department of Physics and Advanced Materials, University of Technology Sydney, PO Box 123, Broadway, NSW 2007, Australia

*Mike.Ford@uts.edu.au

**Abstract.** Alternative materials are required to enhance the efficacy of plasmonic devices. We discuss the optical properties of a number of alloys, doped metals, intermetallics, silicides, metallic glasses and high pressure materials. We conclude that due to the probability of low frequency interband transitions, materials with partially occupied *d*-states perform poorly as plasmonic materials, ruling out many alloys, intermetallics and silicides as viable. The increased probability of electron-electron and electron-phonon scattering rules out many doped and glassy metals.

**Contents**

# 1. Introduction

As nanofabrication techniques become increasingly fast and accurate, the performance of plasmonic systems relies less and less on structure fabrication and more on the fundamental limitations of the underlying materials. Plasmonics has seen an exponential growth, due mainly to the sheer diversity of applications, from optical cloaking [1] to superlensing [melville2005, pendry2000] as well as single molecule surface enhanced raman spectroscopy [2], parasite therapy [3] and optical circuits[4] with high speed optical switching[5, 6].

A number of studies have been performed comparing the plasmonic merit of different metals (see e.g. [7]), and although the alkali metals have increased performance over the noble metals at many frequencies and permittivities, experimental convenience or necessity of inertness dictates that the noble metals are used more frequently.

The free electron character of the alkali and noble metals plays a pivotal role in their plasmonic performance. As the number of free electron metals in the periodic table is severely limited, use of doped metals, alloys and intermetallics to tune the frequency and permittivity response of materials, while simultaneously reducing chemical reactivity and loss, seems an obvious choice.

The concept of loss mitigation by the introduction of gain materials into plasmonic devices is an alternative, though technically more challenging approach. The idea was introduced by Ramakrishna and Pendry [8] who investigated the affect of a gain material replacing the dielectric layers in a multilayer superlensing stack. The imaginary component of the frequency dependent dielectric function $\varepsilon(\omega) = \varepsilon' + i\varepsilon''$, which describes the phase lag of the electrons behind the applied electric field, is negated by a material with a negative effective phase contribution $\varepsilon_{\text{gain}}(\omega) = \varepsilon' - i\varepsilon''$.

The fluorescent dye Rhodamine 6G has been used to compensate for loss in both Local Surface Plasmon (LSP)[9] and propagating Surface Plasmon Polariton (SPP) [9] based systems. Increased confinement of light in metallic waveguides causes a substantial increase in the optical loss. The introduction of gain materials into these systems was proposed by Maier [10]. Quantum dots have been proposed [11] and demonstrated [12] to reduce this loss.

In addition to increased optical loss due to confinement, a number of damping mechanisms contribute to the plasmonic performance of real systems. The effect of grain boundaries in gold films has been discussed quantitatively by Kuttge *et al* where they showed that the characteristic decay length of a propagating SPP mode would be reduced by a grain size dependent factor [13]. Similar effects are apparent in systems with features smaller than the average electron mean free path, and such surface scattering has a significant impact on the optical response of small particles[14] and thin shells (see eg [15]). In addition to surface scattering, the introduction of surfaces allows for the decay of plasmons into electron-hole pairs via surface states (see for example [16] [17]).

It is possible to optimize [18] the size, shape and composition of a multilayer metal system to get the optimum resonance at a particular wavelength, for example to match the absorption profile of a fluorescent dye. Although such systems present yet another way to optimize a plasmonic system, here we shall focus on homogenous systems.

In addition to the conventional methods for tuning the optical response of plasmonic systems, namely modification of structure size and shape; varying the composition of alloyed and intermetallic nanostructures can be used to tune the response. Many studies have investigated the optical properties of noble metal alloy nanoparticles, most notably the Ag-Au alloys, where the plasmon absorption maximum varies linearly from the elemental Ag value of 380nm to the elemental Au value of 520nm as the stoichiometry is varied [19, 20]. This leads one to erroneously assume that the response can be modeled using a simple, linear combination of the experimental dielectric functions. However, this is not the case and a more rigorous description in terms of the movement of the optical gap and Fermi energy is required. Other alloys are also in use in plasmonics, for example Chiu *et al* [21]describe a synthesis technique for producing NiAu alloy nanoparticles in various stoichiometries with the additional property of magnetism. The absorption efficiency of the particles, and hence the plasmon efficacy, reduces with increasing Ni. Other alloys such as the Cu/Zn system show more complex behaviour, with 95% Cu, 5% Zn particles having a higher absorption efficiency than the 100%, 70% and 30% Cu particles[22]. Fabrication of Sn-based intermetallic nanoparticles has resulted[23] in a multitude of shapes including some similar to the familiar split ring resonator often used in meta-materials[24]. Ferrando *et al* have recently published a comprehensive review of nano-alloys and their optical and catalytic properties [25].

Al/Ga and Au/Ga nanocomposites have been used in high speed plasmon polariton modulation [krasavin2006, macdonald2007], and we shall discuss some of the interesting properties of liquid metals in section 7.1.

Although we shall focus mainly on bulk materials, there are particular alloy combinations which do not have stable bulk phases but do however alloy when structured on the nanoscale (see for example FeAg [26]).

To date, reasonably few intermetallic compounds have been used in plasmonics with the main candidates being $AuAl_2$ [27] and $MgB_2$ [kussow2007, limberopoulos2009], with other examples including the Heussler type compound $Co_2FeGa$ [28] and $Au_3Zn$ [29]. Nanoparticles made of the latter show slightly increased absorption efficiency over comparable gold particles.

The transition metal silicides are particularly interesting in light of the ease at which waveguides can be constructed in silicon using conventional semiconductor manufacturing techniques. The transition metals can be masked onto the surface and diffused into silicon, creating guiding structures. In section 5 we discuss their optical properties.

West *et al* have recently discussed the merits of a number of alternative plasmonic systems [30] which we shall not discuss here - namely graphene, semiconductors and phonon polariton materials such as SiC. Here, we present a complimentary review focusing on the electronic properties of some materials not discussed in detail by West *et al*. We largely discount the technical challenges that the utilisation of these materials may impose on device fabrication.

This paper is structured as follows: in Section 2 we shall discuss the necessary electronic features of a good plasmonic material and the properties of some metals. In Sections 3-7 we will discuss the optical properties of other plasmonic materials, which encompass five classes of materials as follows. (3) Alloys - includes mainly non-stoichiometric materials, doped materials and nanograined materials. (4) Intermetallic compounds. (5) Silicides. (6) Systems under pressure. (7) Metallic glasses, liquids and amorphous alloys.

## 2. Plasmonic materials: Definitions and review of metals

*2.1 Quality factors and the ideal plasmonic material*

The complex dielectric function $\varepsilon(\omega) = \varepsilon' + i\varepsilon''$ fully describes the macroscopic electronic response of a material. It is possible to excite a surface plasmon resonance at any frequency for which the real part of the permittivity is less than zero. The quality of the associated resonance depends on the value of the imaginary permittivity at this frequency. For the ideal free electron gas, the dielectric function is usually written in the form of a Drude model:

$$\varepsilon_D(\omega) = 1 - \frac{\omega_p^2}{\omega(\omega + i/\tau)} \quad (1)$$

The plasma frequency, $\omega_p$ is a function of the electron mass and density, and the phenomenological scattering time $\tau$ is sometimes replaced with a scattering frequency $\gamma$ that encompasses all scattering mechanisms including electron-electron, electron-phonon, surface and defect interactions. The bare plasma frequency can be determined by calculating the transition rate at the Fermi surface in the limit of zero energy and momentum transfer [31].

$$\omega_p^2 = \frac{8\pi}{3V} \sum_{i,\mathbf{k}} |\mathbf{P}_{i,i,\mathbf{k}}|^2 \delta(E_{i,\mathbf{k}} - E_F), \quad (2)$$

where P is the momentum matrix element with wavevector **k** within band *i*. This can also be written:

$$\omega_p^2 = \frac{8\pi}{3V} \sum_{i,\mathbf{k}} \left|\frac{dE_{i,\mathbf{k}}}{d\mathbf{k}}\right|^2 \delta(E_{i,\mathbf{k}} - E_F) \quad (3)$$

where $E_{i,k}$ is just the energy of the $i^{th}$ band at wavevector **k**. Thus the bare plasma frequency is just a sum of the gradient of bands at the Fermi surface. In section 6 we will comment on the effect of pressure on the gradient of the bands at the Fermi energy, and in section 3 we shall discuss doping as a method of modifying the Fermi energy.

In real materials, the plasma frequency is shifted from the bare plasma frequency (Eqn 2,3) due to screening by interband transitions, which are single particle excitations from the valence to conduction bands. For example, the screened plasma frequency $\omega_s$ in silver is at 3.8 eV compared to the bare plasma frequency value of 9.6 eV.

We have reviewed and developed a series of metrics to determine the proficiency of metals to perform in particular plasmonics applications[7]. Although every specific geometry will have a different quality factor, in the limit of low loss and the applicability of electrostatics, two generic limiting cases can be derived, (i) for localized surface plasmon (LSP) applications which include the absorption efficiency of nanospheres and nanoshells, and the resolving power of a multilayer and the 'poor mans' superlens and (ii) for extended modes such as surface plasmon polaritons (SPP) and extended LSP modes of ellipsoids.

$$Q_{\text{LSP}} = -\varepsilon'/\varepsilon'', \qquad Q_{\text{SPP}} = \varepsilon'^2/\varepsilon''. \quad (4a,b)$$

In the quasistatic regime, where the features of the plasmonic system are much smaller than the wavelength of light, localized surface plasmons depend on the dielectric function linearly, whereas SPPs depend on the square of the real part. In the limit of low loss, the quality factors can be written in terms of the complex refractive index $m = n + ik$:

$$Q_{\text{LSP}} = k/2n, \qquad Q_{\text{SPP}} = k^3/2n, \quad (5a,b)$$

or the frequency dependent complex optical conductivity $\sigma(\omega) = \sigma_1(\omega) + i\sigma_2(\omega)$:

$$Q_{\text{LSP}} = \frac{\sigma_2 - \omega/4\pi}{\sigma_1}, \qquad Q_{\text{SPP}} = \frac{(\omega - 4\pi\sigma_2)^2}{4\pi\sigma_1\omega}. \qquad (6a,b)$$

By substituting the Drude model (1) into the quality factors (4) and solving for maximum quality we arrive at: [32]

$$Q_{\text{LSP}}^{\max} = \frac{2(\omega_p^2 - \gamma^2)^{3/2}}{3\gamma\omega_p^2\sqrt{3}}, \qquad Q_{\text{SPP}}^{\max} = \frac{\omega_p^2}{2\gamma^2}. \qquad (7)$$

It is now apparent that the most important factor is a large bare plasma frequency to damping ratio $\omega_p/\gamma$. In the event that the scattering rate is unknown, it is often sufficient to make the approximation:

$$\gamma(T) = \varepsilon_0 \omega_p^2 \rho_{\text{DC}}(T), \qquad (8)$$

where $\rho_{\text{DC}}(T)$ is the temperature dependent DC resistivity and $\varepsilon_0$ is the permittivity of free space.

*2.2 A review of the optical properties of metals*

In contrast to gold, the band edge (that is, the frequency at which interband transitions become allowed) in silver is at a frequency above the screened plasma frequency, so that surface modes cannot decay into electron hole-pairs. In gold, the situation is more problematic, since for permittivities $-\varepsilon' < 2$ the surface plasmons decay into electron hole pairs, resulting in a much reduced quality factor at these frequencies.

Unfortunately, the effect of interband transitions is much less localized in frequency than this simple picture portrays. Consider a simple Gaussian distribution representing transitions from some valence to conduction band. The transitions are centered at some frequency $\mu$ with distribution $\sigma$ which approximately describes the dispersion of the bands in the metal, and a number $\alpha$ which describes the number of electrons involved in the transitions (and simultaneously the variation in angular momentum character across the Brillouin zone). The interband spectrum now looks like:

$$\varepsilon_{\text{ib}}''(\omega) = \frac{\alpha}{\sqrt{2\pi\sigma^2}} \exp\left(-\frac{(\omega-\mu)^2}{2\sigma^2}\right) \qquad (9)$$

We now use a Kramers-Kronig integration to determine the real part of the spectrum,

$$\varepsilon_{\text{ib}}'(\omega) = 1 + \frac{2}{\pi} P \int_0^\infty \frac{\Omega \varepsilon_{\text{ib}}''(\Omega)}{\Omega^2 - \omega^2} d\Omega \qquad (10)$$

where $P$ indicates the principal part of the integral, and we set $\omega = 0$:

$$\varepsilon_{\text{ib}}'(0) = 1 + \frac{2}{\pi} \int_0^\infty \frac{\varepsilon_{\text{ib}}''(\Omega)}{\Omega} d\Omega \qquad (11)$$

which can be approximated by $\varepsilon_{\text{ib}}'(0) \approx 1 + 2\alpha/\pi\mu$ if $\mu \gg \sigma$. This of course has the effect that even at frequencies well below the band edge, the real part of the permittivity has additional positive component which degrades the local surface plasmon quality factor:

$$Q_{\text{LSP}}^{\max} = \frac{2(\omega_p^2 - 2\gamma^2[1+\alpha/\pi\mu])^{3/2}}{3\gamma\omega_p^2\sqrt{6(1+\alpha/\pi\mu)}} \qquad (12)$$

This is our second criteria for a high quality plasmonic material: *the number of electrons involved in interband transitions must be low, and at the highest possible frequency.* This simple and quite obvious criteria significantly reduces the number of materials that are likely to have favorable optical properties, by the simple fact that all materials with partially occupied *d* or *f* states are going to perform poorly across the visible due to interband transitions, and even if the transitions do not extend into the IR, poorly at low frequencies because of the aforementioned residual low frequency effect on the polarizability.

The effect of interband transitions on the maximum value of $Q_{SPP}$ is slightly obscure. For low $\varepsilon'_{ib}(0)$, the frequency for maximum $Q_{SPP}$ is just $\gamma$. As $\varepsilon'_{ib}(0)$ increases, maximum $Q_{SPP}$ shifts to lower frequencies. It is apparent from the damping frequencies listed in table 1 that it is not particularly useful to describe the maximum in $Q_{SPP}$. In fact, when the plasma frequency to damping ratio is large, interband transitions have almost no effect on the magnitude of $Q_{SPP}$.

In figure 1, we present the maximum values for $Q_{LSP}$ (bold) and the frequencies at which they occur for all non-group-*f* metals. Due to a combination of high plasma frequency to damping ratios, low probability interband transitions and hence low $\varepsilon'_{ib}(0)$, free electron like metals dominate the periodic table in terms of plasmonic performance. For gold and silver, their quality as plasmonic materials is evident from the sheer number of publications in this area. Reported experimental maximum $Q_{LSP}$ values for silver range from tens [33] to hundreds [34] with the latter matching experimental plasmonic device data more closely [35]. Experimental optical constants for gold show similar variability, with $Q_{LSP}$ values varying between 14 and 34. A number of studies have used the experimental permittivity to compare the plasmonic performance of alkali metals in a number of geometries [7, 36-38]. The Group 13 metals, Ga and In have recently been studied by McMahon *et al* [39] where they report $Q_{LSP}$ values of almost 100 for indium at 3.5 eV using the optical constants of [40], Ga performs better than Sn, Pb, Bi, and Tl over the range 3.5 eV to 12 eV, but still has $Q_{LSP}$ below 10 in this region [39].

The actinides[41] thorium, protactinium[42] and uranium[43] all have interband transitions from *f* to *d* states [41], as do Gd and Dy [44]. These metals perform poorly over most frequencies. For a review of the optical properties of the Lanthanides, see [45].

| Li | Be | | | | | | | | | | | B | C | Max Qlsp Key |
|---|---|---|---|---|---|---|---|---|---|---|---|---|---|---|
| 0.14* | 0.20 | | Element | | | | | | | | | | | 0.00-2.99 |
| **28.82** | **3.58** | | Frequency of Max QLSP | * wmax is at the limit of the available data | | | | | | | | | | 3.00-3.99 |
| Na | Mg | | | | | | | | | | | Al | Si | 4.00-6.99 |
| 1.44 | 4.00 | | Maximum QLSP | # low requency data not included | | | | | | | | 11.00 | | 7.00-9.99 |
| **35.09** | **9.94** | | | | | | | | | | | **13.58** | | 10+ |
| K | Ca | Sc | Ti | V | Cr | Mn | Fe | Co | Ni | Cu | Zn | Ga | Ge | As | Se |
| 1.05 | 0.65* | 0.3* | 0.20 | 0.36 | 0.30 | 0.07* | 0.10* | 0.10* | 0.15 | 1.75 | 3.60# | 8.30 | | | |
| **40.68** | **3.63** | **1.02** | **2.58** | **4.27** | **2.16** | **1.16** | **2.48** | **2.69** | **2.71** | **10.09** | **3.59** | **3.41** | | | |
| Rb | Sr | Y | Zr | Nb | Mo | Tc | Ru | Rh | Pd | Ag | Cd | In | Sn | Sb | Te |
| 0.81 | 0.36* | 1.48* | 3.00 | 0.55 | 0.38 | | 0.10* | 0.30 | 0.10* | 1.14 | 0.65# | 5.10 | 2.25 | 3.50 | |
| **21.90** | **2.85** | **1.41** | **1.16** | **3.39** | **5.38** | | **2.03** | **2.10** | **6.52** | **97.43** | **3.63** | **4.60** | **3.50** | **1.33** | |
| Cs | Ba | Lan | Hf | Ta | W | Re | Os | Ir | Pt | Au | Hg | Tl | Pb | Bi | Po |
| 0.51* | 1.91 | | 0.52* | 0.58 | 0.30 | 0.10* | 0.10* | 0.40 | 0.35 | 1.40 | 4.20 | 3.20 | 5.95 | 3.50 | |
| **11.20** | **0.91** | | **0.79** | **5.25** | **4.96** | **4.99** | **6.12** | **2.55** | **1.96** | **33.99** | **2.20** | **2.71** | **3.07** | **1.15** | |

**Figure 1.** Periodic table of the elements coloured by maximum $Q_{LSP}$. Frequencies are in eV. References are provided in appendix A.

**Table 1.** Optical Constants of metals, including the residual low frequency term caused by interband transitions.

| Element | $\omega_p$ (eV) | $\gamma$ (eV) | $\varepsilon'_{ib}(0)$ |
|---------|-----------------|---------------|------------------------|
| Ag      | 9.60            | 0.0228        | 3.5                    |
| Au      | 8.55            | 0.0184        | 9.6                    |
| Al      | 15.3            | 0.5984        | 10.4                   |
| Na      | 5.71            | 0.0276        | 1.09                   |
| K       | 3.72            | 0.0184        | 1.12                   |

In summary we can characterize the plasmonic performance of materials by considering appropriate ratios of the real to imaginary parts of the permittivity. While the plasma frequency and relaxation time of the Drude model are helpful in this regard, it is also important to consider the shift of permittivity as indicated by the interband component at zero frequency. Considering the best elemental metals, alkalis are free-electron-like but impractical, Au and Ag have significant interband transitions, and Al operates best at very high frequency. These options are restrictive and hence we now consider alternatives to allow more choice of operating frequencies and potentially lower losses.

**3 Alloys**

Nanograined materials can be effectively modeled using a linear combination of the dielectric functions of the constituent metals, weighted by their respective stoichiometries, with an additional damping term to simulate scattering by grain boundaries [46, 47] [48].

*3.1 Noble Metal Alloys*

Due to their interesting electronic structure and colour, the optical properties of noble-noble alloys are among the most studied of metallic compounds. Randomly oriented AuCu, AuAg and AgCu alloys were studied by Rivory [49] both experimentally by evaporation onto glass substrates and using the coherent-potential approximation (CPA) (see [50]). The interband transitions in Au-Ag alloys with silver concentrations of (in atomic %): 0, 21, 41, 62, 94 and 100 were measured using transmittance/reflectance and Kramers-Kronig analysis. The spectra of Au-Cu alloys were made with Cu concentrations of 0, 12, 25, 40, 70, 81 and 100 at. %. Ag-Cu alloys were made with Cu concentrations of 0, 6, 8, 30, 43, 55, 94 and 100 at.% but had to be deposited at 150K to prevent ordering. The crystal size for Au-Ag and Au-Cu are in the range of 300nm to 500nm, whereas for Ag-Cu they are roughly 1.5nm. It is evident from the interband transitions in the Ag-Cu spectra that 1.5 nm grains are large enough to exhibit the effects of short range order, which causes bulk like interband transitions to become evident. A reduction in short range order often indicates a reduction in the strength of interband transitions. This effect is particularly noticeable when studying liquid metals (See for example [51-53] and section 7). Highly ordered Ag-Cu alloys exhibit a similar 2.5 eV $\varepsilon''$ peak magnitude and shift [54]. The interband transition spectra of the three random alloys exhibits a simple 'mixing' where all of the alloy transitions can be attributed to those of elemental gold, silver or copper [49]. Rivory *et al* report the onset of interband transitions in Au-Ag shifts continuously from 2.5 eV in pure gold to 3.9 eV for pure silver, the extremes being in excellent agreement with the data of Kreibig [55]. A comparison with the optical data of Johnson and Christy[34] and Weaver *et al*[56] show excellent agreement with the elemental data of Rivory *et al*. For Au-Cu the onset of interband transitions shifts from 2.5 eV for pure gold down to an apparent minima at 2.2 eV for 50 at.% Au where it stays for Au concentrations down to pure copper. Part of this shift is due to a decrease in the lattice constant [57]. In all three alloys, the metal with larger interband transitions dominates the spectra. The order of transition strength for these three metals is Cu > Au > Ag.

Doping of Cu with Al has been shown to introduce indirect transitions near the L point, and the secondary band edge shifts to lower energies, while the primary band edge generally gets larger with increasing dopant concentration [58]. This effect also occurs for Cu doped with Ga, Zn, Sn, Si and Ge [59].

The random binary alloy Cu-Fe was made by Korn *et al* in various iron concentrations from 0 at.% to 20 at.% [60]. The elemental copper peak at approximately 5 eV does not shift a great deal, indicating that the Fermi level does not shift very much upon alloying, but the transition strength decreases, indicating a reduction in order. This effect occurs in Cu-Mn and to an extent in Cu-Pd and Ag-Pd as well [61]. However, in none of these alloys is the minimum in $\varepsilon''$ less than that of the constituent elements.

Alloying of Ag with Mg and Cd causes an additional peak to appear below the 3.9 eV band edge of elemental silver [62, 63], which itself is reduced and allows for the excitation of bulk plasmons above 5 eV, albeit with reduced efficacy due to the overlap with interband transitions, similar to gold. The addition of Sn causes a dramatic increase in the scattering rate up to 0.6 eV (similar to bulk aluminum), and no additional interband transitions are visible, the band edge seems to maintain the same magnitude and energy as that of elemental silver. The relaxation time increases marginally for the Mg and Cd alloys, with maximum values of approximately 0.2 eV and 0.15 eV respectively. They report $\gamma = 0.04\,\text{eV}$ for elemental silver.

Silver-indium alloys show the interesting property that the transitions that make up the 3.9 eV peak can be shifted upon alloying with indium [64]. The 3.87 eV $L_3 \rightarrow L_2$ transition shifts to higher energies and the 4.03 eV $L_2 \rightarrow L_1$ shifts to lower energies. The magnitude of the 3.9 eV peak decreases with increasing In concentration, but the plasma frequency decreases and the damping frequency increases. The overall effect of alloying on $\varepsilon''$ is to increase the imaginary permittivity, resulting in a minimum of 1.5 at 3.5 eV for 12 at.% indium compared to the minimum value for silver, where $\varepsilon'' = 0.37$. The introduction of Ni defects in gold has a similar effect, increasing the damping frequency due to impurity scattering, but reduces the magnitude of the band edge [65].

*3.2 Transition Metal alloys*

Thomas and Thurm performed optical experiments on binary alloys of W, Ta, Re and Ir in various stoichiometries [66]. Although the optical properties of the alloys are not linearly dependent on the optical properties of the elements, the magnitude of the interband transition maximum in $\varepsilon''$ always lies in between the maximum values for the elements. The maximum in the interband spectrum shifts as follows through the alloy series: W: 20 at 1.8 eV, Re: 16 at 2.5 eV, Ta: 14 at 3 eV.

The prospects for alloys in general can be summarized by observing that although they allow tuning of the real part of the permittivity, they typically have interband strengths similar to their constituents. Grain size effects in some alloys (e.g. AgCu) may modulate interband transitions at the expense of degraded relaxation time due to scattering in a similar fashion to amorphous materials.

**4. Intermetallics**

In this section we shall discuss the optical properties of the intermetallic compounds in order of the number of constituent elements. As compounds they allow tuning, but avoid grain issues and can have totally different properties compared to the reactants. We survey alkali-metal, noble-noble, group 13 binaries, other binaries and some ternary compounds.

*4.1 Alkali Metal Binary Intermetallics*

Optical measurements of LiAl in the NaTl structure suggest that this compound is an excellent free electron metal with only a small interband transition around 0.55 eV, [67] whereas resistivity measurements present conflicting data, with values around 10 μΩcm [68, 69] compared to silver and gold with values of 1.77 μΩcm and 2.66 μΩcm respectively [70].

We showed recently [71] that some of the alkali noble intermetallics had optical gap to plasma frequency ratios greater than 1, indicating that it was very unlikely that interband transitions would disrupt the optical response of these materials. Unfortunately, calculations of the DC resistivity indicated that the Drude phenomenological scattering rate was too high for these compounds to compete with silver and gold [32]. A comparison of the local surface plasmon quality factors for these materials, alongside experimental data for silver and gold is presented in figure 2.

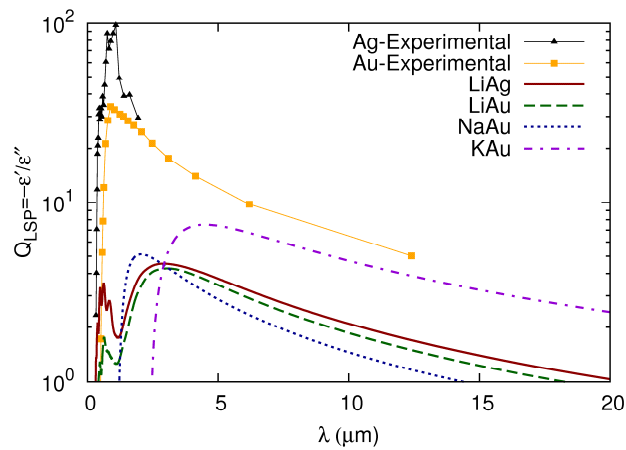

**Figure 2.** Comparison of calculated local surface plasmon quality factors for the alkali-noble binary intermetallics with the most favourable experimental values for silver [72] and gold [73].

*4.2 Noble Binary Intermetallics*

Rivory *et al* also investigated the effect of short range ordering on the stoichiometric alloy $AuCu_3$. With increasing order a new peak appears at approximately 3.6 eV. Slightly different results were reported by Scott *et al*; electropolishing of samples post annealing was shown to cause the peak to appear at 3.28 eV [74] but Skriver and Lengkeek noted that electropolishing preferentially etched grains in their polycrystalline sample and reported the peak at 3.6 eV [75]. They also remove the intraband contribution from their experimental data using a Drude fit and note two additional peaks: one at 0.8 eV and another at 1.2 eV. $CoPt_3$ and $MnPt_3$ crystallise in the same structure as $AuCu_3$ but partially occupied *d*-states result in many low energy transition mechanisms, resulting in $Q_{LSP} < 1$ between 1.5 eV and 5.0 eV [76].

The noble - group III alloys crystallize in the $CaF_2$ structure. The noble atoms occupy sites on an FCC cell and the group III atoms form a simple cubic structure in the centre of the FCC cell [77]. The discovery of the purple coloured, gold aluminium alloy AuAl$_2$ is often attributed to Sir Roberts-Austen [78]. It has since received a great deal of attention, not only due to its colour and applications in jewellery, but also for its possible applications as an energy efficient window coating. Cortie *et al* [27] measure and calculate the reflectance spectra of AuAl$_2$ using density functional theory. They show an experimentally determined reflectance minima at 2.5 eV, which has been shown to persist for Al:Au ratios of between 3.2:1 to 1:1 [79]. Minor discrepancies appear between the position of the measured and calculated reflectance minima due to self interaction errors. They also measure the reflectance of PtAl$_2$ films which show a reflectance maximum at around 1.9 eV of 55% which steadily decreases into the infrared.

Vishnubhatla *et al* [80] studied the optical properties of AuAl$_2$, AuGa$_2$ and AuIn$_2$. They note that interband transitions appear at 2.2 eV in AuAl$_2$ and are responsible for the reflectance minimum at 2.5 eV. Hsu *et al* note that this transition is not due to Au 5d bands [77]. The onset of interband transitions decreases as the atomic number of the alloying compound increases. AuGa$_2$ has much broader experimental interband transitions in the region around 2 eV than AuAl$_2$ [80, 81]. Calculations show that this broadening is caused by an additional transition at approximately 1.6 eV. When substituting platinum for gold the reflectance minima at 2 eV shifts to 3eV and the reflectance peak at 3 eV shifts to 4 eV. This occurs due to a combination of shift in the 2 eV transition peak to 3.6eV in PtGa2, and an associated shift in the real part of the permittivity. The variation in the position of the transition peak between gold and platinum can be explained by a decrease in energy of the 5d bands of Au in AuGa$_2$ causing the $\Gamma_7$ band to be below the Fermi energy [81]. This effect is not seen in AuAl$_2$ or AuIn$_2$ [hsu1994, switendick1969].

Silver indium alloys show the interesting property that the transitions that make up the 3.9 eV peak in silver can be shifted upon alloying with indium [64]. The result is that the 3.87 eV L$_3 \rightarrow$ L$'_2$ transition shifts to higher energies and the 4.03 eV L$'_2 \rightarrow$ L$_1$ shifts to lower energies. Unfortunately, the overall effect is to increase the imaginary permittivity, resulting in a reported minimum of 1.5 at 3.5 eV for 12 at.% indium. The minimum $\varepsilon''$ for silver occurs at approximately 3.5 eV with a value of 0.22. The addition of aluminium into silver also causes a continuous shift in the minimum of $\varepsilon''$ from 0.22 to a value of 1.78 at 2.7 eV [82]. This increase in the minimum of $\varepsilon''$ causes a decrease in the magnitude of the real part of the permittivity, causing the main silver transition at 4 eV to shift to higher energies and its value to shift from 1.75 to just above zero, causing the permittivity of the alloy to become negative for all energies below 7.6 eV [82].

Plasma frequencies for the three CsCl structured Noble alloys AuZn, CuZn and PdIn are shifted quite substantially from their unscreened locations at 11.5 eV, 12.6 eV and 11.0 eV respectively by interband transitions [83]. The biggest shift occurs for PdIn, where the screened plasma frequency is shifted to 2.4 eV. In AuZn the screened plasma frequency $\omega_s$ is shifted the least with $\varepsilon'$ crossing 0 at 7 eV. The real permittivity for CuZn exhibits a positive region between 2.5 eV and 3.1 eV, giving it a purplish pink colour similar to that of AuZn which is yellow-pink. The interband transition strength is greatest in CuZn and least in PdIn. The onset of transitions appears to occur at approximately 1eV for PdIn and AuZn, and at about 2 eV for CuZn. All three compounds exhibit a transition gap between 2eV and 3eV in the region where $\varepsilon'$ becomes steep due to intraband contributions. As the amount of disorder in CuZn is increased the band edge at 2eV reduces and a low energy transition become apparent [84].

Of all the alloys studied here, CsCl structured binary intermetallics show the most promise, with large optical gaps and in some cases, plasma frequencies comparable to silver and gold.

*4.3 Group 13 Binary Intermetallics*

NiAl, CoAl and FeAl all have major interband contributions to the imaginary part of the permittivity, with the minimum value of $\varepsilon''$ being 10 for CoAl and FeAl at 2.75 eV and 3.25 eV respectively [85]. Of Ni$_3$Al [86], Ni$_3$Ga and Ni$_3$In, only Ni$_3$In shows reasonable $Q_{\text{LSP}}$, with values above 2 at frequencies below 1 eV [87]. DFT calculations on Ni$_3$Al and CoAl both suffer from the effect of partially occupied *d*-states, and exhibit the property that electron self energy corrections shift the interband transitions to lower energies, and hence Rhee *et al* use a negative lambda fitting regime, which corrects their optical spectra into line with experiment.

The optical properties of CoAl and NiAl alloys were determined by Kim *et al* [88] both experimentally and using the linearized augmented plane wave (LAPW) method. Their calculations show very good agreement with experiment below 2.5eV, and accurately describe the onset of interband transitions which occurs at 1eV for CoAl and 2.5 eV for NiAl. The variation in the onset of interband transitions is attributed to a shift in the Fermi energy with the addition of a *d*-electron in nickel. In CoAl, transitions across a band pair along Γ-M -X and X-R are responsible for the low energy transitions. When nickel is substituted for cobalt, the top band becomes occupied and transitions become impossible. The minimum $\varepsilon''$ they report for NiAl is approximately 11 at the low energy edge of their experimental data (1.2 eV) and approximately 4.5 at the high energy edge of their experimental data (6.0 eV) for CoAl. The FP-LAPW method using GGA (PW92 functional) and self energy corrected LDA (LDA + U) exchange correlation functionals were used by Rhee *et al* [89] to calculate the optical properties of FeAl. It was found that both LDA+U and a positive λ fitting routine were required to improve the agreement with experiment as the effects of correlation are known to induce a paramagnetic ground state in FeAl. The experimental spectra exhibits three main transitions at 0.5 eV with magnitude 102, one at 1.5 eV with magnitude 29 and the last at 4 eV with magnitude 12. Transitions below 3 eV occur around the Γ, X and M special points [90]. The minimum in $\varepsilon''$ occurs at the high energy edge of the experimental data at 5 eV with a value of 9.6. The calculations indicate that the transitions decrease over this region. Other groups [91] report maximum transition magnitudes of 55, almost half that of Rhee *et al*.

In 1985 van der Heide *et al* [86] performed ellipsometric spectroscopy on two Ni$_3$Al alloys. One of the samples was an 8mm polycrystalline sample and the other a 3mm single crystal. The sample size limited the accuracy of the results they obtained at energies greater than approximately 5.5eV and hence their measurements extend only from 0.5eV to 5.5eV. They report no noticeable difference between the optical measurements of the two samples.

The most interesting feature in the spectra is a peak in $\varepsilon''$ from about 2.5eV to about 4.5eV. This peak pushes the real part of the permittivity above 0 resulting in two bulk plasmons, one at 3.4eV and another at 3.9eV. An analysis of the infrared Drude tail showed that in the absence of interband transitions the sample would exhibit a bulk plasmon at approximately 9eV (which is quite close to the bulk plasmon of nickel at 10 eV [73]). Unfortunately due to a high damping frequency, the Drude contribution to $\varepsilon''$ grows rapidly at visible wavelengths. Van der Heide *et al* also calculated that interband transitions appear at the optical conductivity from approximately 1eV in Ni$_3$Al and claim that early onset interband transitions are common in *d*-metals. A comparison of the optical conductivity with the Joint Density of States (JDOS) per energy level showed remarkable similarity in their results. This is particularly surprising as they leave the transition matrix elements constant. This means that the heights of the peaks that are evident in the JDOS/ω are arbitrary, even when compared across the spectrum.

Further investigations by Rhee *et al* showed an increase in interband transitions with non-stoichiometric concentrations as well as increased transition strength with ferromagnetism and temperature [rhee1997, rhee2003]. Rhee *et al* also note that the transitions at 4 eV originate from bands near M along Σ rather than around Γ as reported by [86]. A reduced transition strength is reported by Hsu *et al* [87], which causes the real part of the permittivity to be positive for energies above 2.4 eV. They attribute this to a superior sample surface.

Hsu and Wang [87] recently calculated the optical properties of the alloys $Ni_3Al$, $Ni_3Ga$ and $Ni_3In$ using DFT with a FP-LAPW basis. They calculated interband transitions between 0eV and 150eV however no $\varepsilon'$ results are presented. In similar fashion to the lattice constants reported for the group-13 gold alloys, the $Ni_3Al$ and $Ni_3Ga$ lattice constants are quite similar at 3.571 and 3.589 and the $Ni_3In$ lattice constant is 3.745 about 5% greater than $Ni_3Al$. The optical calculations overestimate the magnitude of both the 1.0 eV and 2.0 eV transitions in $Ni_3Al$, the 1.25 eV transition in $Ni_3Ga$ and all transitions in $Ni_3In$. Notably, the minimum in the experimental $\varepsilon''$ for $Ni_3In$ occurs at 1.35 eV and has a value of 0.5; at this energy the real part has value of -3.5, giving a $Q_{LSP}$ of 7.

The Laves phase ($MgCu_2$ structure) Lanthanide group 13 intermetallics $LaAl_2$ $CeAl_2$ $PrAl_2$ $YbAl_2$ all have *f*-states very close to the Fermi energy. Interaction with conduction states results in a low energy threshold for interband transitions [lange2000, lee2000]. However, the compounds $YAl_2$ $LuAl_2$ have their *f*-states centered well above and well below the Fermi energy respectively, resulting in the main interband transition occurring at approximately 2 eV. None of these materials have $Q_{LSP}$>1. The $AuCu_3$ compound $LuAl_3$ also has high lying *f*-states, and the density of states at the Fermi energy of $YbAl_3$ is nearly twice as large compared to $ScLa_3$ and $LuAl_3$ because of partially occupied *f*-states, but interband transitions cause the compound to have no metallic character between 1.5 eV and 5.5 eV [92].

*4.4 Other Binary Intermetallics*

In $LaSn_3$ the onset of interband transitions occurs at approximately 0.5 eV with the main peak at approximately 1.5 eV [93]. All the transitions between 1 eV and 5 eV can be explained due to the mixing of 4*f* character into states near the Fermi energy, with the dominant transition mechanism arising from La 5*d* to hybrid *f-d-p* states [93]. In $CeSn_3$ the situation is even worse due to increasing 4*f* character of conduction states, which causes partial occupation of *f-d-p* states allowing for additional transition mechanisms. $ThPd_3$ and $UPd_3$ both exhibit the $TiNi_3$ structure, $UPd_3$ having partially occupied 5*f*-states and $ThPd_3$ having completely unoccupied 5*f*-states; the onset of interband transitions in both materials is < 1 eV [94]. The interband component of the optical spectra for all these materials is too great to allow for reasonable plasmonic activity.

*4.5 Ternary Intermetallics*

MgAuSn exhibits the cubic AlLiSi structure (F43m) and is coloured purple due mainly to very strong interband transitions around 3 eV [95]. The transitions are likely due to the parallel band effect which gives aluminium its strong interband component at 1.5 eV. Because the transition in MgAuSn becomes steep at a higher energy, and the material has a lower effective Drude plasma frequency, Kramers-Kronig integration forces the real part of the permittivity into the positive over the region 2.2 eV to 3.1 eV. Interband transitions were calculated using the tight-binding linear muffin orbital (TB-LMTO) method within the local density approximation. Intraband contributions were included by fitting a Drude damping constant and plasma frequency to experimental data.

The optical properties of the magnetic Heusler alloy $Cu_2MnAl$ have been calculated using a tight binding plane wave [96]. Generally reasonable accord between experiment and theory was apparent below 3.5 eV which included the two main features at 1.5 eV and 2.7 eV. The feature at 1.5 eV also bears similarity to the low energy transition in aluminium, and the 2.7 eV peak is reported to arise from transitions between hybridised conduction bands. Additional studies by Kudryavtsev *et al* [97] suggest that $Q_{LSP}$ is greater than 1 for frequencies below 1 eV. Damping to plasma frequency ratios of between 8.85 and 10.7 have been measured in samples of $Ni_2MnGa$ depending on the annealing temperature [98] however both $Ni_2MnGa$ and $Ni_2MnIn$ [99] have large low energy interband transitions. $Fe_2TiAl$ has a plasma frequency of 0.22 eV [100], lower than the value of 1.32 eV for $Fe_2VGa$, which gives a reasonable $\omega_p/\gamma$ value of 29 [101], however interband transitions cover most of the spectrum, disrupting quality.

The most promising of all the alloys studied here is $Li_2AgIn$ alloy. It crystallizes in the NaTl-type structure (Zintl Phase), with 16 atoms per unit cell such that a group of 8 BCC cells make up a cube. The corner atoms of each BCC cell are composed of alternating lithium and indium atoms and the centers of alternating BCC cells are lithium and silver atoms. Zwilling *et al* [102] report some very interesting results for this compound, the most surprising of which is that at exactly 2 eV the imaginary part of the permittivity is zero. A material such as this would have phenomenal optical properties, most notably, an infinitely sharp resonance. Of course, such a resonance is not really possible, and 'very large' will have to suffice. The real part of the permittivity at 2 eV is also reported by Zwilling *et al* [102] and a value of -16 is given. Zwilling *et al* determined the complex permittivity by ellipsometry. Their samples were prepared by melting the constituent metals in various ratios in a furnace at 1000◦ C. The concentration of lithium remained constant while the silver and indium concentrations obeyed the formula $Ag_{2-x}In_x$. At $x = 1$, $\varepsilon''$ in $Li_2AgIn$ is at a minimum of 0.0 at 2.0 eV. In $Li_2Ag_{0.50}In_{1.50}$ the minimum in $\varepsilon''$ increases to above 4, shifts to lower energies and the gradient of the real part becomes less negative. Unfortunately, in the same work, Zwilling *et al* report the optical properties of $Li_2CdIn$. According to their data, the imaginary part of the permittivity reaches -2 at approximately -3.1 eV. This violates causality and cannot be correct. It is possible that the cause for this error lies in the extrapolation regime they use in conjunction with the Kramers-Kronig relations, but no reference is made. Overall, this indicates that the proximity of $\varepsilon''$ to zero for $Li_2AgIn$ should be discounted, but not ignored.

We recently [71] calculated the interband component of the imaginary permittivity for a series of materials in the series $Alkali_2$-Noble-Group 13 and Alkali-$Noble_2$-Group 13, with the alkali metals consisting of Li, Na and K, the noble metals Ag and Au and the group 13 metals Al, Ga and In. Figure 3 compares the imaginary permittivity measured by Zwilling *et al* with calculated interband imaginary permittivity values of our previous work [71] no Drude intraband term is included.

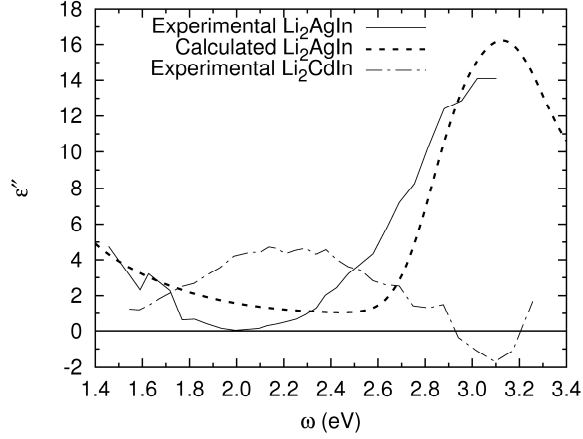

**Figure 3.** Experimental imaginary permittivity of Zwilling *et al* [102] for $Li_2AgIn$ (solid line) and $Li_2AgCd$ (dash dot line). Note the non causal region between 2.9 eV and 3.2 eV. Calculated interband imaginary permittivity of Blaber *et al* [71] for $Li_2AgIn$ (dashed line) shows the permittivity not going to zero.

In conclusion, the plasmonic quality of intermetallic compounds is heavily reliant on the complexity of the band structure. Interband transitions dominate in materials with large numbers of atoms in the unit cell. If alternative plasmonic materials are going to be realized in the form of intermetallics, binary compounds with only two atoms in the primitive cell that have low lying *d*-states are most likely to be competitive.

## 5. Silicides

Silicides present an interesting alternative to metals for use in plasmonics. Due to extensive use in metal-oxide-semiconductor field effect transistors (MOSFETs) their optical and electrical characteristics have been widely studied (for a review of their application to MOSFETs see [103]). Moreover, as most of the studied silicides grow epitaxially on silicon, current and future semiconductor manufacturing techniques can be directly applied to the creation of sub-10nm plasmonic devices. In fact device structures amenable to on chip plasmonic information transfer such as nanowires have already been created from silicides [jiang2009, song2007, zhang2006]. Recently, Soref *et al* [104] have proposed the use of $Pd_2Si$ as an alternative to gold in SPP based plasmonic devices. They argue that as long as on chip transport requirements are met (i.e. high speed, small size, low loss) then the operating wavelength of the system is irrelevant. To ensure that the mode confinement is reasonable the longest operating wavelength is chosen such that field emanating from the waveguide penetrates no greater than three wavelengths on the air exposed side. Unfortunately, they mistakenly divide all the penetration lengths by a factor of $2\pi$, causing the operating wavelength to be at very low frequencies, where in fact the penetration length reaches hundreds of microns. Nonetheless, their criterion is a good one and we shall use it to assess the plasmonic performance of a number of silicides.

The penetration depth of the tangential component of a surface plasmon polariton traveling along a metallic strip surrounded by air can be approximated by:

$$L_{pen} = \frac{\lambda}{\text{Re}[\sqrt{-1/(\varepsilon(\omega)+1)}]} \qquad (13)$$

where $\lambda = c/\omega$. Enforcing the condition $L_{pen}/\lambda = 3$ for a Drude metal reduces to $\varepsilon'(\omega) \approx -10$. The propagation length of such a mode is:

$$L_{\text{prop}} = \frac{1}{2\,\text{Im}[\lambda\sqrt{\varepsilon(\omega)/(\varepsilon(\omega)+1)}]}. \tag{14}$$

Where the frequency is (for a Drude metal):

$$\omega = \sqrt{\frac{\gamma^2(1-\varepsilon') - \omega_p^2}{\varepsilon'-1}} \tag{15}$$

A summary of the optical properties of some silicides is presented in table 2. We have included the SPP quality factor at the telecommunications wavelength of $1.5\,\mu m$. For comparison, the value for silver is 4522 [30]. We previously noted (see Section 1.2) that a Drude model is sufficient to describe $Q_{\text{SPP}}^{\max}$ as long as the plasma frequency is much larger than the scattering rate - this is of course arises from a significant contribution to the real part of the permittivity that drowns out the effect of interband additions to the imaginary part and any residual effect on the polarizibility that these transitions may incur. However, all the silicides presented in table 2 have small plasma frequency to damping ratios compared to silver and gold. The data for $TiSi_2$ is derived from the plasma frequency and DC resistivity, and although most of the scattering rates calculated from DC resistivity measurements are very close to those extracted from optical data, there is no a priori way of estimating the error. More recent optical constants by Kudryavtsev et al [105] for $TiSi_2$ reduces $Q_{\text{SPP}}$ at 1.5 um to 27, down from 1348.

**Table 2.** Optical constants of various transition metal silicides including surface plasmon polariton quality factors at the telecommunications wavelength $\lambda\sim 1.5\mu m$. Calculated using the collated data of Nava et al [106]. $\gamma_{\text{opt}}$ is the Drude scattering rate extracted from optical constants, and is used - along with the plasma frequency - in the calculation of $Q_{\text{SPP}}$. In the case where $\gamma_{\text{opt}}$ was not available, $\gamma_\rho$ - the scattering rate calculated from the DC resistivity – was used instead. $Q_{\text{SPP}}^{\max}$ corresponds to $Q_{\text{SPP}}$ at the damping frequency (see section 2.1).

|  | $\omega_p$ (eV) | $\gamma_{\text{opt}}$ (eV) | $\gamma_\rho$ (eV) | $Q_{\text{SPP}}$ (0.8 eV) | $Q_{\text{SPP}}^{\max}$ | $L_{\text{prop}}$ ($\mu m$) when $L_{\text{pen}}/\lambda = 3$ |
|---|---|---|---|---|---|---|
| $VSi_2$ | 2.75 | 0.070 | 0.069 | 112 | 769 | 61.64 |
| $NbSi_2$ | 2.3 | 0.051 | 0.052 | 100 | 1030 | 9.59 |
| $TaSi_2$ | 2.6 | 0.060 | 0.055 | 115 | 942 | 64.45 |
| $NiSi_2$ | 4.6 | 0.150 | 0.157 | 160 | 471 | 80.89 |
| $NiSi$ | 3.8 | 0.035 | 0.035 | 470 | 5887 | 234.80 |
| $Ni_3Si$ | 3.4 | 0.044 | 0.149 | 293 | 2999 | 150.03 |
| $V_5Si_3$ | 2.9 | 0.143 | 0.149 | 60 | 203 | 33.82 |
| $V_3Si$ | 3.4 | 0.115 | 0.114 | 109 | 431 | 57.27 |
| $HfSi_2$ | 1.5 | - | 0.026 | 55 | 1620 | 48.70 |
| $GdSi_2$ | 2.4 | - | 0.091 | 62 | 348 | 36.40 |
| $ErSi_2$ | 1.3 | - | 0.014 | 60 | 4540 | 70.55 |
| $TiSi_2$ | 4.2 | - | 0.015 | 1348* | 38240 | 661.11 |
| $WSi_2$ | 1.78 | 0.020 | 0.004 | 127 | 4003 | 90.72 |
| $Pd_2Si$ | 2.8 | 0.030 | - | 276 | 4376 | 149.20 |

Doping of transitions metals exhibits similar qualities to doped metals, namely, the screening of interband transitions. The $Fe_{1-x}Si_x$ compound studied by Kim et al [107] (figure 4) shows decreasing metallicity with increasing silicon concentration as the density of conduction electrons decreases, this causes a decrease in the SPP properties of FeSi. However, increasing silicon concentration also screens interband transitions, and for some frequencies and compositions, the LSP quality is greater than that of pure iron.

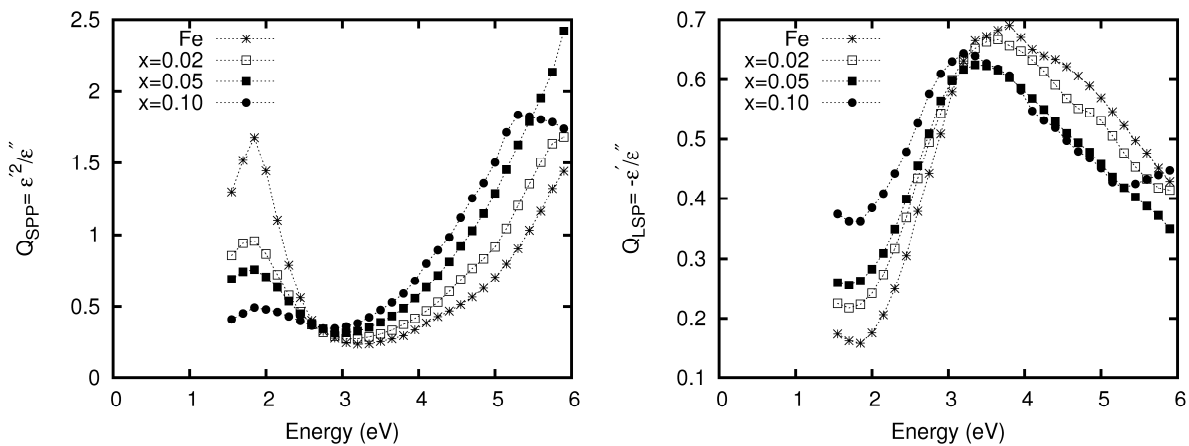

**Figure 4.** Quality factors for surface plasmons in the structure $Fe_{1-x}Si_x$

In figure 5 we present experimental $Q_{SPP}$ data collated by Nava et al [106] and compare it to the Drude model for which optical constants are presented in table 2. Due to the distinct lack of agreement between experimental optical constants and the Drude model for these compounds, or the requirement of frequency-dependent scattering rates, we strongly discourage its use, even at very long wavelengths.

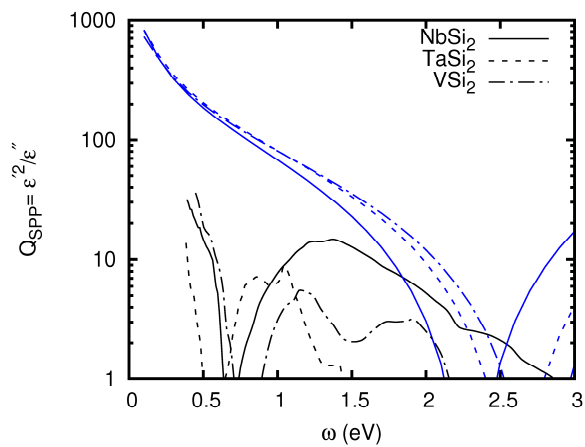

**Figure 5.** Local surface plasmon quality factors for group 5 transition metal silicides. Black lines: experimental data collated by [106], blue lines: Drude model using parameters from table 2.

The alkali metal silicides have been shown to be semiconductors [108]. The alkaline earth metal silicides exhibit poor metal characteristics (eg CaSi $\rho_{DC}$ =282 μΩcm [109]) semiconductors eg $BaSi_2$ [110] and reasonable metallic character (eg $CaSi_2$ $\rho_{DC}$ =32 μΩcm [109]). Some of the rare earth silicides are metallic, such as GdSi [111] and ErSi [wetzei1991, angot1999] albeit with exceptionally poor plasma frequency to damping ratios ($\omega_p/\gamma \approx 1$) whereas others are semiconductors such as EuSi and YbSi [112].

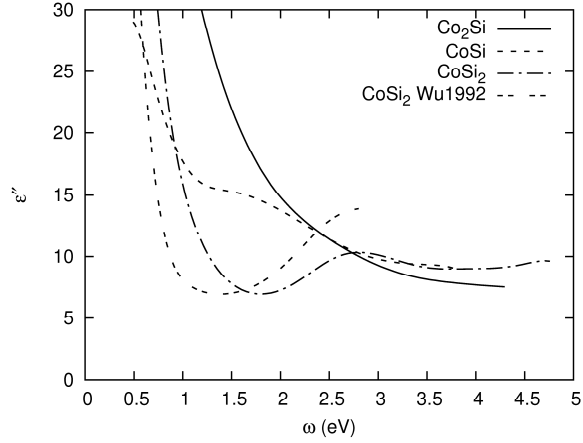

**Figure 6.** Experimental imaginary permittivity of cobalt silicides in various stoichiometries. Data from Kudryavtsev *et al* [113] and Wu [114].

In summary, although the silicides offer considerable technological conveniences and some of them are potential candidates for guiding long wavelengths (e.g. $TiSi_2$), they are generally poor when compared to other materials and are not suitable for LSP applications.

## 6. High Pressure Materials

Metals and alloys under pressure has become a very popular topic. Pressure and phase transformations drastically alter the electronic properties of materials, shifting optical gaps, the plasma frequency as well as modifying electron-phonon and electron-electron scattering.

Structural phase transitions, electronic topology transitions and metal-insulator transitions in systems under high pressure open the doors to whole new world of materials for plasmonics. We shall briefly discuss some of these properties, with a focus on pressures sustainable in active devices.

Potentially, an increase in pressure can increase the plasma frequency meeting one of our criteria for a good plasmonic material. However, increasing the band gradient at the Fermi surface, and the possibility of overhauling the topology of the Fermi surface due to pressure induced electron transfer from one band to another can dramatically alter the electron phonon coupling, and hence have a detrimental impact on the phenomenological relaxation time. This change in topology, where previously unoccupied bands cross the Fermi energy is known as an Electronic Topology Transition (ETT). Such transitions can have positive and negative effects. The effects of pressure on the alkali metals has become an exciting topic due to the discovery of novel phase transitions[115], some of which are superconducting (see e.g. [116] and [117]).

The band edge in lithium can be shifted from 3 eV to almost 7 eV by the application of pressures as low as 40 GPa [118]. The magnitude of the band edge is increased from 1.75 in the BCC phase to approximately 2.75 for the FCC and hR1 phases. The FCC phase has a screened plasma frequency to optical gap ratio greater than 1[119].

The introduction of pressure on K causes a more substantial increase in the interband contribution to the permittivity; $\varepsilon''_{ib}(\omega_g)$ increases from less than 1 at 2.2 eV to greater than 7 when the crystal volume is reduced to 45% of the ground state volume [120]. Gao *et al* have studied an anomaly in the resistivity of SC calcium[121]. They note that the resistance anomaly in SC calcium that occurs at approximately 40 GPa can be attributed to an increase in the electron phonon matrix element and an increase in the plasma frequency due to an ETT from 4*s* to 3*d* states. The pressure dependence of the plasma frequency for some alkali, noble and group 13 elements is shown in figure 7, although we can expect an increase in plasmonic quality with increased plasma frequency, the technical difficulty of operating a plasmonic system under pressure negates any (minor) increases in quality. Additionally, in many cases additional pressure increases resistivity [122].

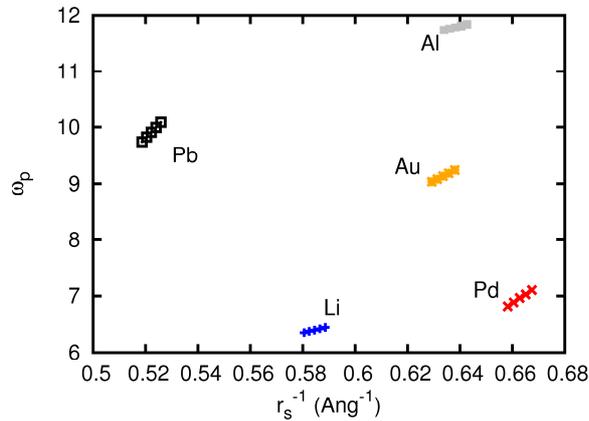

**Figure 7.** The effect of pressure on the plasma frequency of some metals studied by Sundqvist [123]. $r_s$ is the effective radius for an atom in the unit cell. Pressure increases to the right.

ETTs in the Noble-Group 13 intermetallics $AuX_2$ under pressure cause additional transition mechanisms at the high symmetry points in the Brillouin zone, presumably reducing the resonance quality [124].

In summary, although topological changes under high-pressure offer the chance to drastically alter material properties, and some of them may exhibit increased plasmon quality (e.g. in Li), others are degraded (K). Further, the pressure required is impractical (although not impossible).

## 7. Liquid Metals and Glassy/Amorphous Materials

For any purely amorphous system, the concept of k-space is ill-defined, and as such the difference between interband and intraband transitions disappears resulting in a smearing of the band edge. Usually, the smearing occurs down to zero frequency as transitions which are forbidden in the crystalline state become possible with increasing amorphicity. As such, the optical properties of such systems should be well described by the Drude model with a high damping frequency. There are some serious disadvantages to using liquid metals and amorphous alloys in plasmonics, notably, the generally high temperature required to have a metal in a liquid state, and the difficulty in depositing nano-patterned amorphous films. Nonetheless, we shall see that liquid sodium has quite amenable optical properties, and the amorphous silicide PdSi may be useful in plasmonic devices.

*7.1 Liquid Metals*

The reduced order and increased temperature evident in liquid metals has the effect of shifting the interband transitions to lower energies as they are broadened in a similar fashion to amorphous compounds with the additional drawback of increased electron-electron scattering rate due to increased temperature. Although one would not expect interband transitions in the optical spectra of liquid metals due to disruption to the periodic potential, even short range order in liquid metals is sufficient to observe interband transitions.

The quality factors $Q_{LSP}$ and $Q_{SPP}$ are presented for liquid and solid Na in figure 8a using the data of Inagaki *et al* [inagaki1976liquid, inagaki1976solid]. Of all the materials in this review, liquid sodium is the first material to exhibit superior plasmonic properties compared to its standard state elemental counterpart. Ingaki *et al* measure the optical constants under a pressure of $2 \times 10^{-9}$ Torr and at a constant temperature of 120ºC to minimize vaporization. Once the sodium was melted, an oxide coating on the sample was removed in situ by the use of a stainless steel scraper. The Maximum $Q_{LSP}$ of liquid Na is 42.1 [51], whereas for solid Na it is 41.1, and the frequency at which this maximum occurs increases in liquid Na to 2.3 eV from 1.3 eV in the solid. From these parameters for liquid Na we can calculate the scattering rate and plasma frequency for a Drude metal: $\gamma = 0.038$ eV and $\omega_p = 4.16$ eV. There are remnants of the band edge evident in the liquid phase [51], albeit with much reduced magnitude, and a slight shift to higher energies. Helman and Baltensperger [125] argue that this apparent interband component can be explained by frequency dependent scattering due to ion-electron interactions.

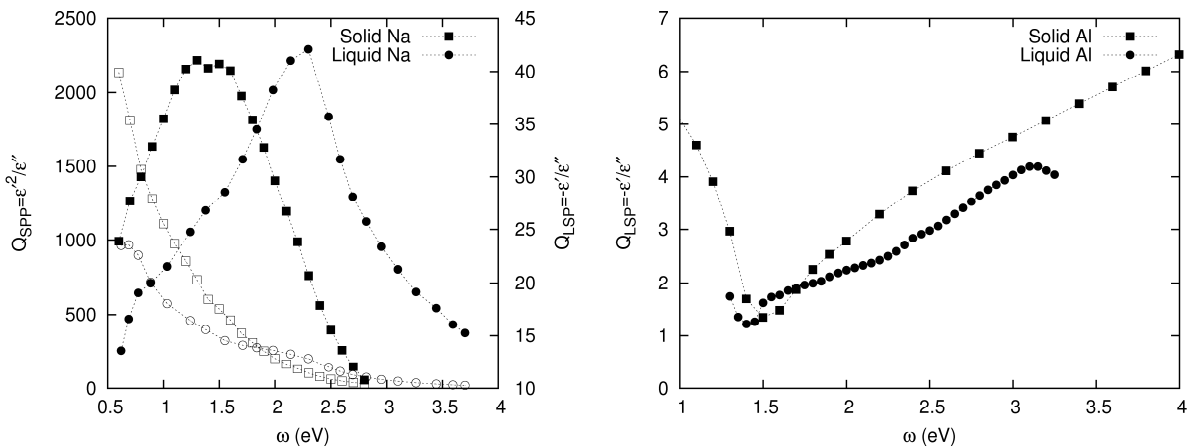

**Figure 8.** Optical data for A) sodium [inagaki1976liquid, inagaki1976solid] and B) aluminum[126, 127]. Data for solid phase (squares) and liquid phase (circles). Filled symbols represent $Q_{LSP}$ data and open symbols represent $Q_{SPP}$ data.

$Q_{LSP}$ data by Krishnan and Nordine [126] for liquid Al is presented in figure 6b alongside the data for the solid state by Shiles *et al* [127]. The 1.5 eV transition is shifted to lower energies, and there is a noticeable increase in the scattering rate which is not evident in liquid Na. Additionally, the plasma frequency reduces with increasing temperature [128].

Al-Ga and Au-Ga nanocomposites have been employed in high speed modulation of surface plasmon polaritons. The functionality arises from a heating effect in the composite that causes a structural transformation in the gallium [6, 129, 130]. A comparison of local and propagating plasmon modes for liquid and solid gallium is presented in figure 9. Liquid gallium has a significantly superior optical response over the solid phase for a very large wavelength range (100 nm to 20 µm). The maximum $Q_{SPP}$ is 185 at 1.77 µm.

The band edge in liquid silver shifts to higher energies, the onset of interband transitions is not as steep in the liquid phase, and the maximum of the band edge is at 4.5 eV. In copper, the band edge broadens and shifts to lower energies. Both materials have increased scattering rates [131].

The scattering rate in liquid Pb and Sn is an order of magnitude greater than in the solid state [132]. The maximum $Q_{LSP}$ measured for liquid Pb is 1.38 at 3.7 eV, more than 4 times lower than the solid [133]. Similarly, Bi has maximum $Q_{LSP}$ of 0.7, 5 times lower than the solid [133].

Mercury and liquid mercury-indium alloys were studied by Hodgson [134]. The scattering rate increases from 1.44 eV to 1.72 eV when the temperature of elemental mercury is increased from 20 C to 200 C. Adding indium to liquid mercury steadily decreases the scattering rate at 20 C and can reach values of 0.8 eV with 33.4 at.% indium.

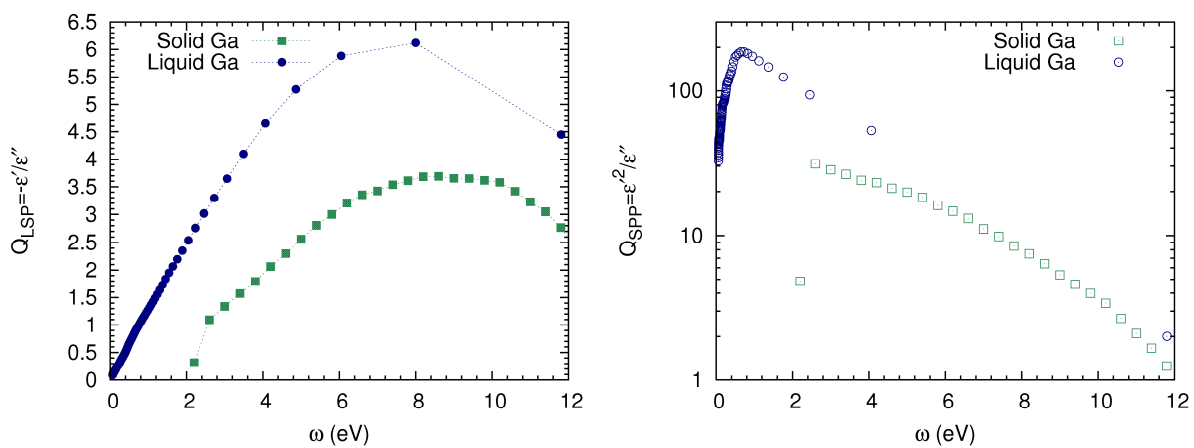

**Figure 9.** Optical data for solid [135] and liquid [136] gallium A) $Q_{LSP}$ data B) $Q_{SPP}$ data.

Both silicon and germanium exhibit the free electron like Drude tail in the liquid phase at the moderately impractical temperatures of 1600K and 1300K. Fuchs [137] has compared experimental and calculated data for liquid silicon prepared in a variety of ways, and some of the data shows $Q_{LSP}$ can take values greater than 1 at energies above 4 eV.

*7.2 Amorphous/Glassy Alloys*

Amorphous $Au_{1-x}Si_x$ [138, 139] was the first material discovered to have a damping frequency greater than the optical gap, resulting in exceptionally poor optical properties. The summary of the maximum $Q_{LSP}$ and the frequency at which it occurs is presented in figure 10.

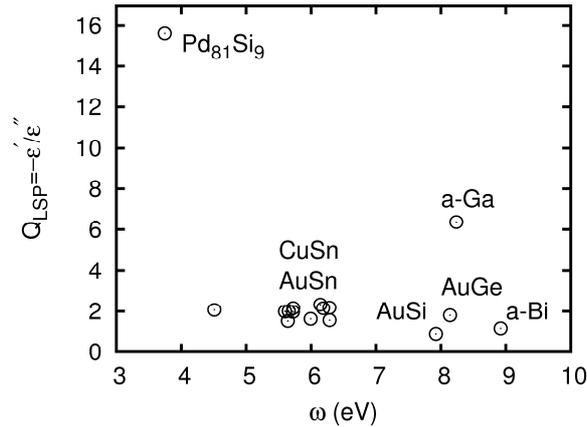

**Figure 10.** Maximum $Q_{LSP}$ for a variety of metallic glasses using plasma frequencies and drude damping parameters collated by Mitzutani [140]. The materials have varying stoichiometries defined by $A_xB_{100-x}$. The AuSi [138] and AuGe have $x$=75 and $x$=70.

With the exception of amorphous PdSi and amorphous Ga, a majority of the glasses reviewed by Mitzutani [140] in figure 10 have poor optical properties.

The optical constants of the amorphous alloys of $Ag_xSn_{1-x}$ and $Au_xSn_{1-x}$ were measured at a single wavelength by Loistl and Baumann [141]. They approximate the intraband damping parameter and show that it is greatest at $x$=0.5 for AuSn and $x$=0.66 for AgSn, reaching values of 2.1 eV in both cases, indicating some sort of additional order at these stoichiometries where chemical bonding begins to occur. Similar effects occur in liquid CsAu, MbBi and LiBi (see [142]).

Amorphous NiP alloys were studied by McKnight *et al* [143]. They show that in the amorphous phase interband transitions are shifted to higher energies with increasing phosphorous content. Unfortunately, although the scattering rate reduces with increasing P, the plasma frequency also decreases. The relationship between the square of the plasma frequency and the scattering rate is linear for P concentrations below about 20 % indicating additional scattering mechanisms begin to appear at these concentrations. In CoP [144] any alloying with phosphorous causes the low energy transition in Co to become unresolvable.

Amorphous Ti and Mo disilicides have inferior plasmonic properties compared to the crystalline compounds studied in section 5 [105]. Amorphous Fe and Cr have a smaller interband component than the crystalline versions, but Kudryavtsev *et al* [105] report no data for the real part of the permittivity.

Summarizing, amorphous and liquid materials exhibit smeared interbands which reduces their impact. However, since most metals require higher temperature to exist as a liquid, electron-electron scattering is typically increased. For most amorphous and liquid materials we surveyed plasmonic performance is worse, however liquid phase Na and Ga exhibit increased performance over their solid state counterparts.

## 8. Conclusions

We have discussed the optical performance of some different types of materials - including the elemental metals, alloys, intermetallics, silicides – as well as high-pressure and amorphous phases. The silicides and many of the alloys studied here have partially occupied *d*-bands that adversely affect their plasmonic performance. However, dopants may be introduced to disrupt the low energy transitions, giving an overall increase in plasmonic quality. The silicides as well as most liquid and amorphous materials and doped materials have very large Drude phenomenological scattering rates. We showed that the plasma frequency and band edge of some materials could be shifted substantially with the addition of pressure, however, plasmonic devices working under 40 GPa of pressure is a scientific curiosity at best, and practical applications are difficult to envision. We note the very interesting optical properties of liquid sodium and gallium, and would be very interested in the optical properties of the NaK eutectic, which has a melting point of -30°C.

We conclude that intermetallic compounds are most likely to offer an alternative to silver and gold for plasmonic applications. Materials with simple crystal structures and low lying d-states, are most likely to perform well.

**Appendix A**

**Table A.1**) References for the optical properties of the elements presented in Figure 1. In the case where multiple tabulations are available (in particular for Cu, Ag and Au), we chose the optical constants that gave the highest $Q_{LSP}$. The data is partially sourced from collations by Weaver [73] and Palik [45, 145]. Unpublished data by Weaver and coworkers (V, Hf, Re, Os) made available in [73], is cited as such. Data for Zr is not cited correctly by Weaver and Frederikse [73].

| Elements 3-28 | | Elements 29-49 | | Elements 50-83 | |
|---|---|---|---|---|---|
| Li | [146] | Cu | [33] | Sn | [147] |
| Be | [148] | Zn | [149] | Cs | [150] |
| Na | [151, 152] | Ga | [135] | Ba | [153] |
| Mg | [33] | Rb | [150] | Hf | [73] |
| Al | [127] | Sr | [153] | Ta | [154] |
| K | [152, 155] | Y | [156] | W | [56] |
| Ca | [149] | Zr | [73] | Re | [73] |
| Sc | [157] | Nb | [158] | Os* | [73] |
| Ti | [159] | Mo | [154] | Ir | [160] |
| V | [73] | Ru | [73] | Pt | [161] |
| Cr | [162] | Rh | [160] | Au | [73] |
| Mn | [159, 163] | Pd | [164] | Hg | [165] |
| Fe | [166] | Ag | [33] | Tl | [167] |
| Co | [166] | Cd | [149] | Pb | [168] |
| Ni | [169] | In | [40] | Bi | [170] |

**Acknowledgements**

The authors are grateful for useful discussions with Paul West, Vladimir Shalaev, Alexander Kildishev and Chris Poulton. This work was supported by the Australian Research Council, and the University of Technology, Sydney.